\documentclass{aa}  

\newcommand\dosingle[1]{#1}  \newcommand\dodouble[1]{ } 

\newcommand\nice[1]{#1}    \newcommand\subm[1]{}   
\subm{ \renewcommand\dosingle[1]{ }   \renewcommand\dodouble[1]{#1} }

\dodouble{ \documentclass[referee]{aa} }

\usepackage{graphicx}

\usepackage{amsfonts}
\usepackage{url}

\newcommand\prerefereechanges[1]{#1}    

\newcommand\postrefereechanges[1]{#1}

 \usepackage{hyperref}

\nice{ 
\providecommand{\eprint}[1]{\href{http://arxiv.org/abs/#1}{{\tt [arXiv:#1]}}}

\providecommand{\url}[1]{\href{#1}{#1}}

}

\providecommand{\adsurl}[1]{} 
\newcommand\SSS{Sect.~}
\providecommand\PRL{PRL}
\providecommand\jetpL{JETPLett}
\providecommand\jrasc{J. Roy. Ast. Soc. Canada} 

\nice{ \newcommand\mycaptionfont{\protect\footnotesize} }

\newcommand\gtapprox{\,\lower.6ex\hbox{$\buildrel >\over \sim$} \, }
\newcommand\ltapprox{\,\lower.6ex\hbox{$\buildrel <\over \sim$} \, }
\newcommand\propapprox{\,\lower.6ex\hbox{$\buildrel \propto\over \sim$} \, }

\newcommand\arcs{\ifmmode {'' }\else $'' $\fi}     
\newcommand\arcm{\ifmmode {' }\else $' $\fi}       
\newcommand\ddeg{\ifmmode^\circ\else$^\circ$\fi}    

\newcommand\frtoday{Le\space\number\day\space\ifcase\month\or
  janvier\or f\'evrier\or mars\or avril\or mai\or juin\or
  juillet\or ao\^ut\or septembre\or octobre\or novembre\or 
d\'ecembre\fi\space \number\year}
\newcommand\todayISO{\number\year-\ifnum\month<10 0\fi\number\month-\ifnum\day<10 0\fi\number\day}

\newcommand\cqg{ClassQuantGra}   %
\newcommand\BASI{Bull. Astr. Soc. India}

\newcommand\hGpc{\mbox{$h^{-1}$ Gpc}}

\newcommand\rinj{{r}_{\mathrm{inj}}}  
\newcommand\rSLS{r_{\mathrm{SLS}}}  
\newcommand\zSLS{z_{\mathrm{SLS}}}  
\newcommand\rC{R_{\mathrm{C}}}  

\newcommand\Omm{\Omega_{\mathrm{m}}}

\newcommand\ximc{\xi_{\mathrm{C}}} 
\newcommand\xisc{\xi_{\mathrm{A}}} 

\newcommand\Npoint{N_{\mathrm{p}}}

\providecommand\tablefootmarkmath[1]{^{#1}\,\ignorespaces}

\title{The size of the Universe according to the Poincar\'e dodecahedral space hypothesis}

\author{Boudewijn F. Roukema 
  \and
  Tomasz A. Kazimierczak 
}

\institute{Toru\'n Centre for Astronomy, Nicolaus Copernicus University,
ul. Gagarina 11, 87-100 Toru\'n, Poland 
}

\date{\frtoday}

\titlerunning{Poincar\'e space Universe size}
\authorrunning{Roukema \& Kazimierczak}

\begin{document}


\newcommand\Nchainsmain{16}
\newcommand\Npergroup{four}

\abstract
{One of the Friedmann-Lema\^{\i}tre-Robertson-Walker (FLRW) models that
  best fits the Wilkinson Microwave Anisotropy Probe (WMAP) sky maps
  of the cosmic microwave background (CMB) is that whose comoving
  space is the Poincar\'e dodecahedral space. The optimal fit of this
  model to WMAP data was recently found using
  an optimal cross-correlation method.  For geometrical reasons, there
  was concern that systematic error in the estimate of the
  matched-circle (observer-centred) angular radius $\alpha$, or
  equivalently, the (comoving) size of the Universe $2\rinj$ (twice
  the injectivity radius), might be much higher than the random error.}
{In order to increase the falsifiability of the model,
  especially by multiple imaging of collapsed objects, it would be
  useful to reduce the uncertainty in this estimate and to estimate
  the fraction of the sky where multiply imaged gravitationally bound
  objects should potentially be detectable.}
{A 
    corollary of the matched circles principle---{{\em the existence of
      matched discs}}---is introduced in order to describe a useful
    subset of multiply imaged objects. The cross-correlation method at
    $\ltapprox 1$~{\hGpc} is applied to WMAP 7-year data in order to
    improve the estimate of $\alpha$.}
{The
  improved matched-circle radius estimate is
        $\alpha = 23 \pm 1.4\ddeg$,
  where the uncertainty represents 
  systematic error dependent on the 
  choices of galactic mask and
  all-sky map.
  This is equivalent to $2\rinj = 18.2\pm 0.5$~{\hGpc} for
  matter density parameter $\Omm=0.28\pm 0.02$. The lowest
  redshift of multiply imaged objects is 
  $z=106\pm18$. Multiply imaged
  high overdensity (rare) peaks visible during $200>z>106$  should 
  be present in matched discs of radius 
  $14.8\pm2.3^\circ$.
}
{The accuracy in the matched circle radius estimate is considerably
  improved by using the higher resolution signal. The predicted
  matched discs (over $200>z>106$)
  project to about 20\% of the full sky. Since any
  object located exactly in the discs would be multiply imaged at
  equal redshifts, evolutionary effects would be small for objects
  that are nearly located in the discs.}


\maketitle

\dodouble{ \clearpage } 


\newcommand\talphabest{
\begin{table}
\caption{\mycaptionfont 
Matched-circle radius $\alpha$ at which the Gpc-scale cross-correlation
\prerefereechanges{$\bar\xi_{\beta}(\alpha)$} is maximised.
\label{t-alphabest}}
$$\begin{array}{l cccc} \hline
\rule{0ex}{2.5ex}
\mbox{map} & \multicolumn{4}{c}{\beta} \\
  & 0.033 & 0.1 & 0.2 & 0.4 \\ 
\hline
\rule{0ex}{2.5ex}
 \mbox{ILC, 85}\tablefootmarkmath{a}  & 23.0 & 23.0 & 23.0 & 23.0\\
 \mbox{V, 85}\tablefootmarkmath{a}  & 21.8 & 23.0 & 24.4 & 24.4\\
 \mbox{W, 85}\tablefootmarkmath{a}  & 22.0 & 22.4 & 22.4 & 34.4\\
 \mbox{ILC, 75}\tablefootmarkmath{b}  & 23.8 & 23.8 & 23.8 & 23.8\\
\hline
\end{array} $$
\tablefoot{
\tablefoottext{a}{KQ85 mask}
\tablefoottext{b}{KQ75 mask}
}
\end{table}
}  

\newcommand\fzminalpha{
\begin{figure}
\centering 
\includegraphics[width=70mm,bb=78 173 489 580]{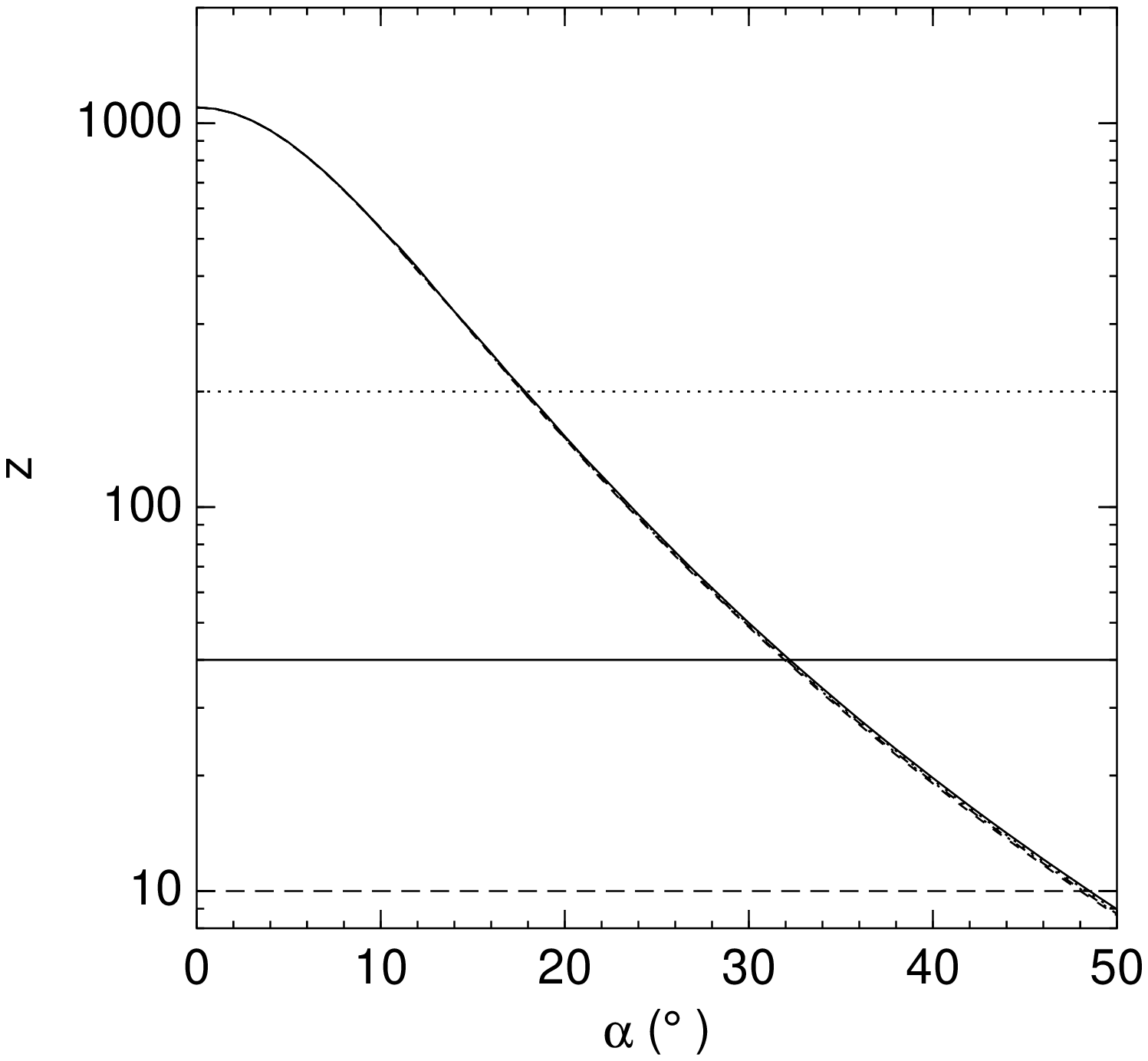}
\includegraphics[width=70mm,bb=78 173 489 580]{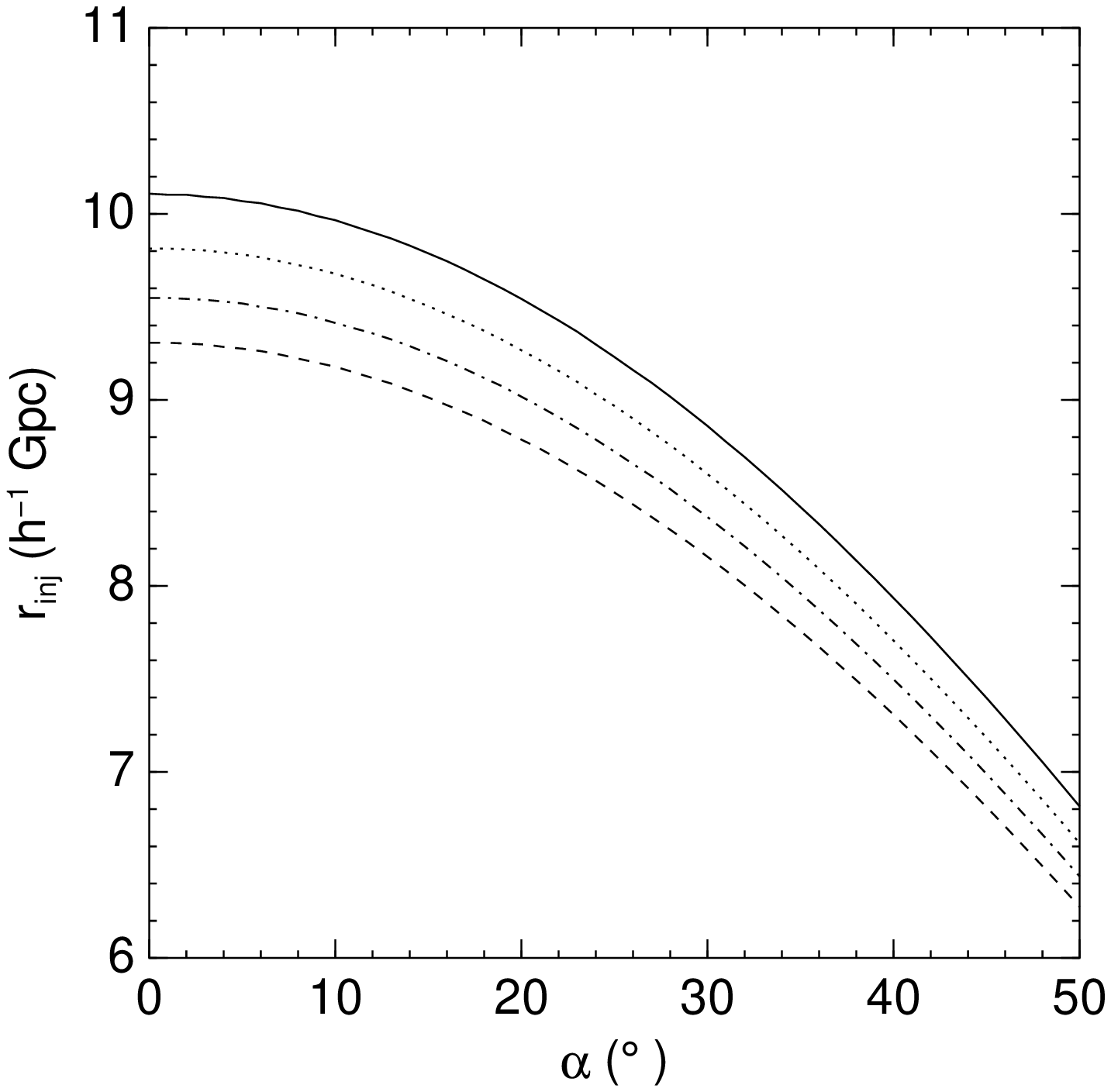}
\caption[]{ \mycaptionfont {\em Upper panel:} 
  \prerefereechanges{Redshift $z_{\min}$ at}
  the centre of a matched disc in a Poincar\'e space as a function of
  matched circle radius $\alpha$, for $\Omm=0.26, 0.28, 0.30, 0.32$ as
  curved solid, dotted, dot-dashed and dashed lines, respectively
  (almost indistinguishable). \prerefereechanges{Redshifts 
    $z_{\min}=10, 40,$} and $200$ are indicated
  as illustrations of epochs of successively rarer expected formation
  of galaxies with Population III stars.  {\em Lower panel:}
  Corresponding injectivity radius of the Universe $\rinj$.}
\label{f-zminalpha}
\end{figure} 
} 

\newcommand\fxiilcmain{
\begin{figure}
\centering 
\includegraphics[width=8cm]{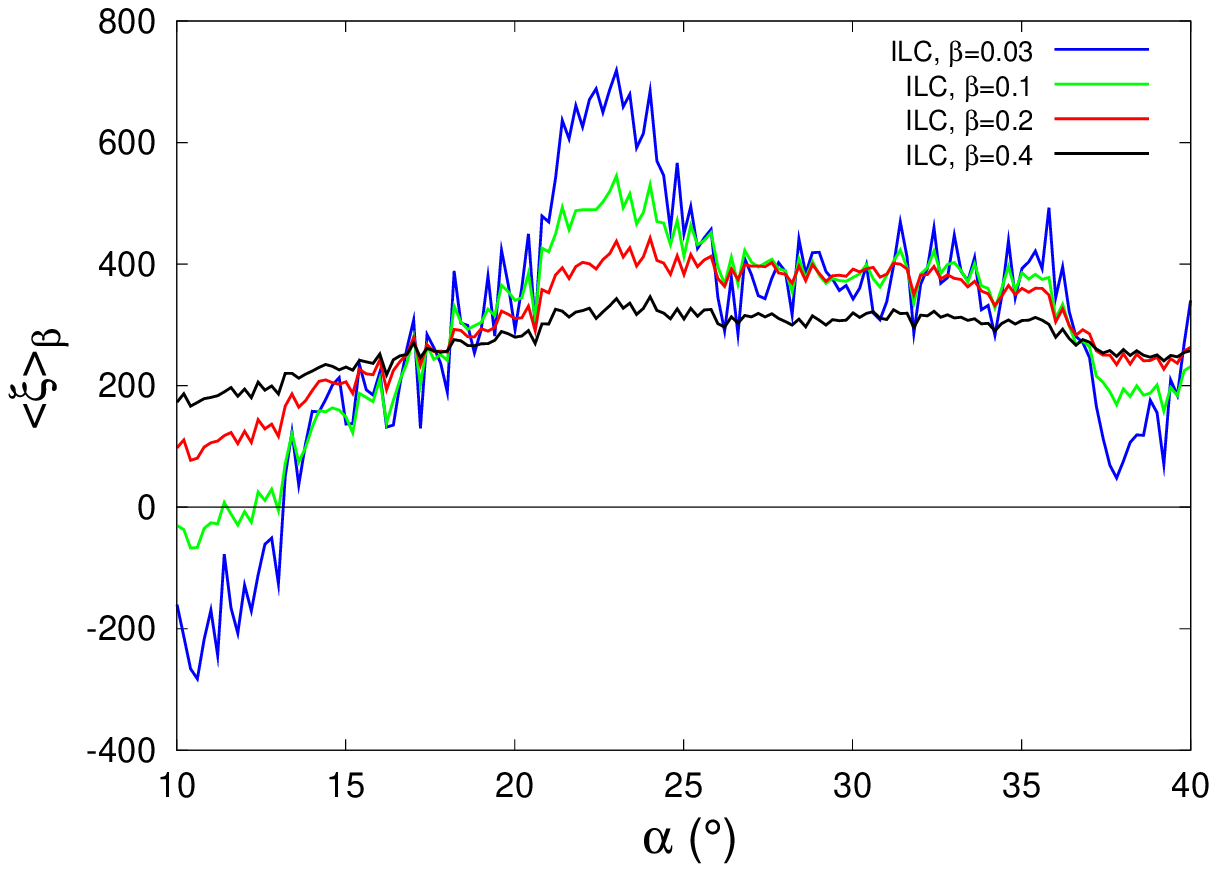}
\caption[]{ \mycaptionfont 
Gigaparsec-scale cross-correlation 
\prerefereechanges{$\bar\xi_{\beta}(\alpha)$} [Eq.~(\protect\ref{e-defn-xim})] 
\prerefereechanges{in $\mu \mathrm{K}^2$}
as a function of matched-circle radius
$\alpha$, for the 
WMAP 7-year 
ILC map, for fractions of the  surface of last scattering radius
$\beta = 0.03, 0.1, 0.2, 0.4$ as indicated (from top to bottom at
$\alpha \sim 23^\circ$), using the KQ85 galactic mask.
}
\label{f-xi-ilc-85}
\end{figure} 
} 

\newcommand\fxiV{
\begin{figure}
\centering 
\includegraphics[width=8cm]{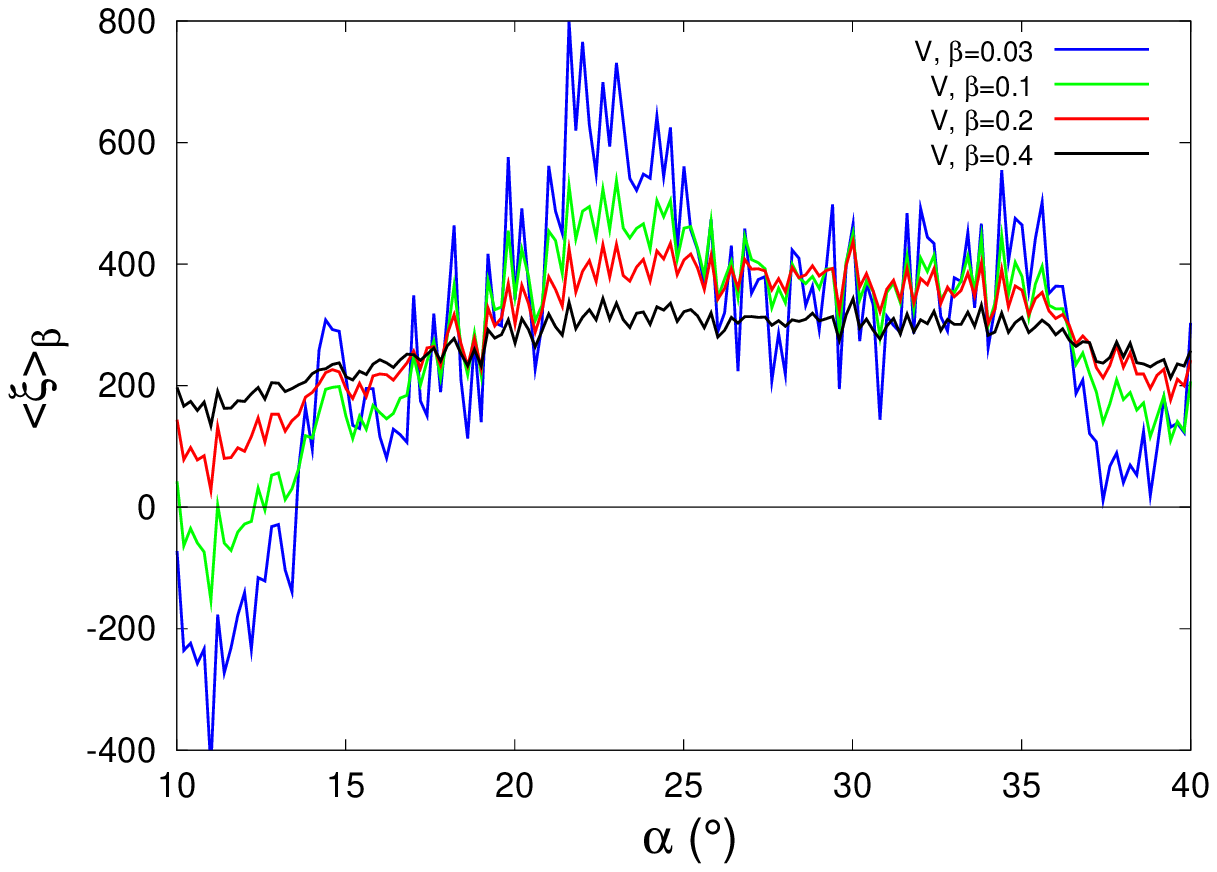}
\caption[]{ \mycaptionfont 
Gigaparsec-scale cross-correlation 
\prerefereechanges{$\bar\xi_{\beta}(\alpha)$}, as for 
Fig.~\protect\ref{f-xi-ilc-85}, 
for the WMAP7 foreground-corrected V map.
}
\label{f-xi-V-85}
\end{figure} 
} 

\newcommand\fxiW{
\begin{figure}
\centering 
\includegraphics[width=8cm]{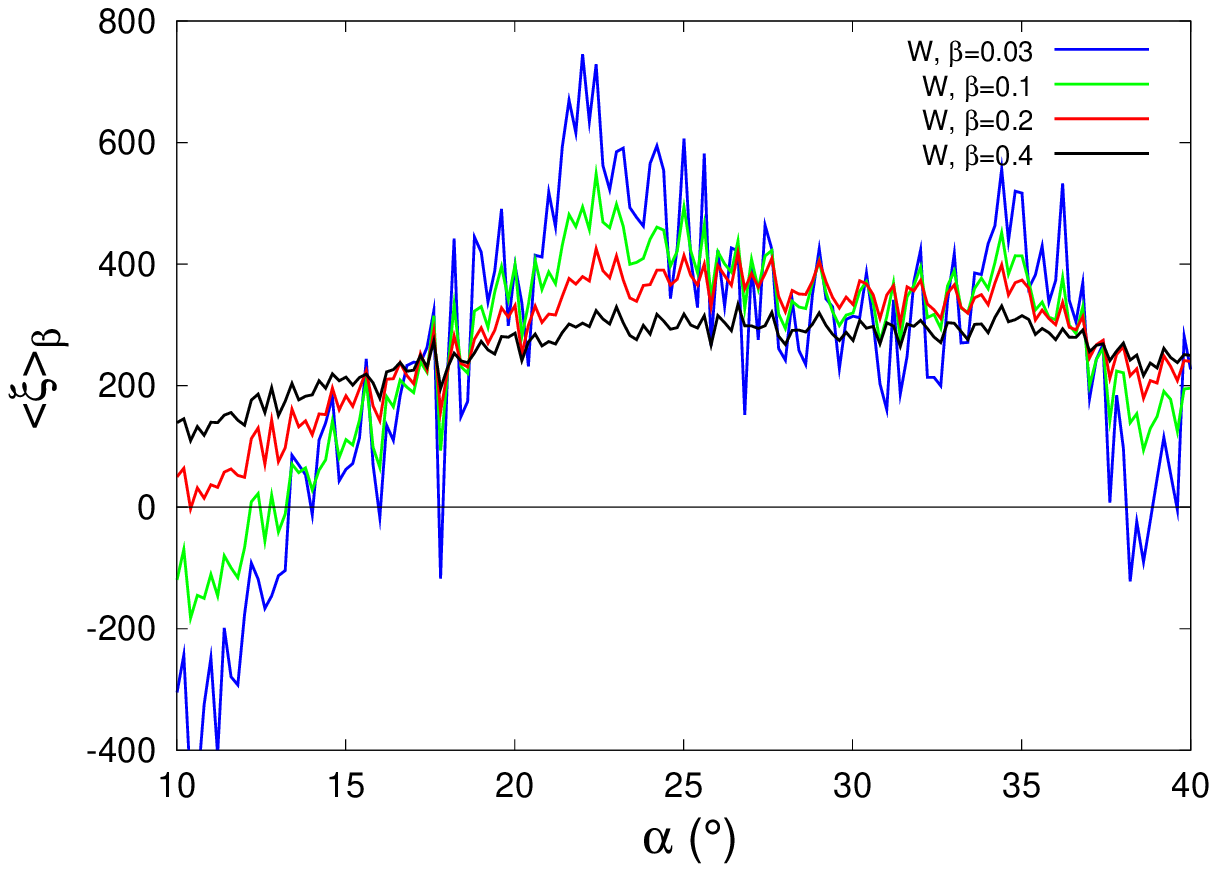}
\caption[]{ \mycaptionfont 
Gigaparsec-scale cross-correlation 
\prerefereechanges{$\bar\xi_{\beta}(\alpha)$}, as for 
Fig.~\protect\ref{f-xi-ilc-85}, 
for the WMAP7 foreground-corrected W map.
}
\label{f-xi-W-85}
\end{figure} 
} 

\newcommand\fxiilcalt{
\begin{figure}
\centering 
\includegraphics[width=8cm]{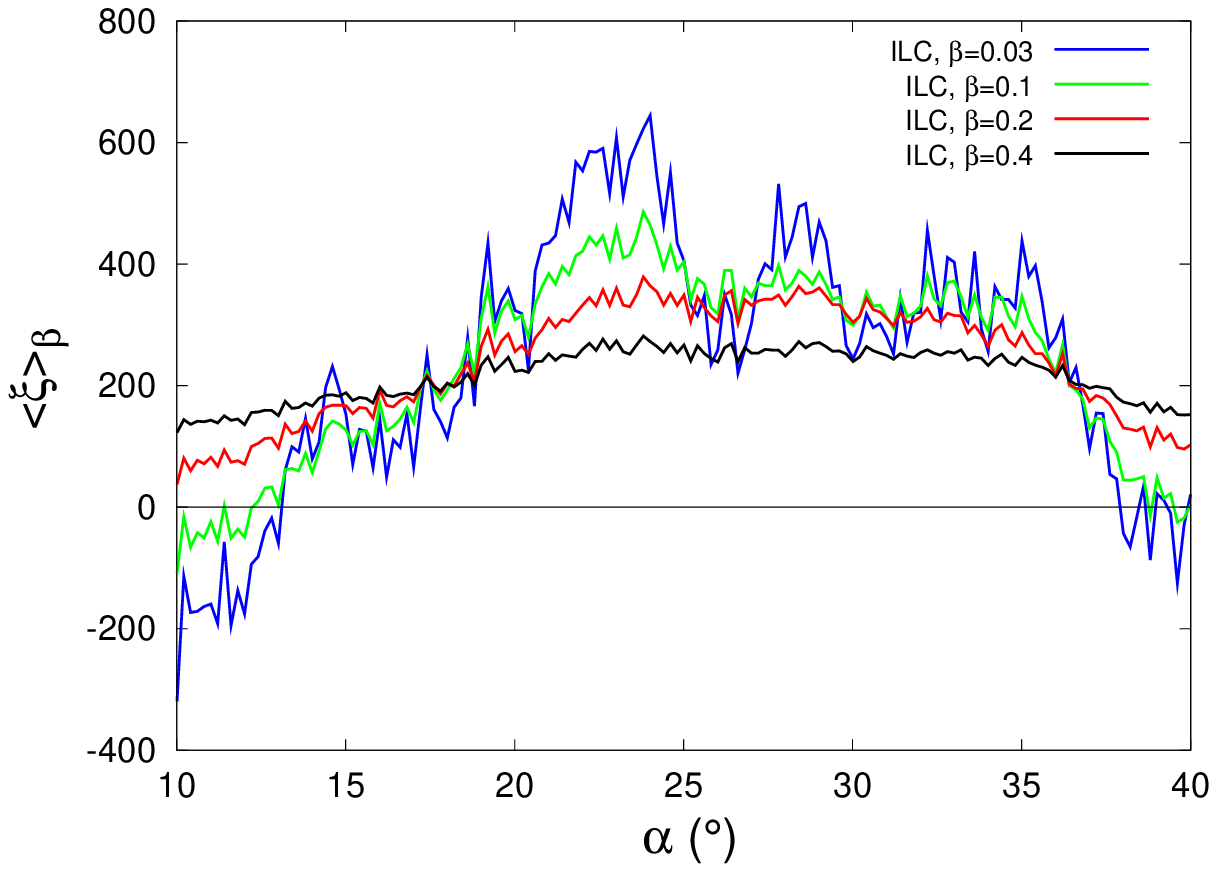}
\caption[]{ \mycaptionfont 
Gigaparsec-scale cross-correlation 
\prerefereechanges{$\bar\xi_{\beta}(\alpha)$}, as for 
Fig.~\protect\ref{f-xi-ilc-85}, 
for the WMAP7 ILC map and the KQ75 galactic mask.
}
\label{f-xi-ilc-75}
\end{figure} 
} 

\newcommand\fcorreg{
\begin{figure}
\centering 
\includegraphics[width=8cm]{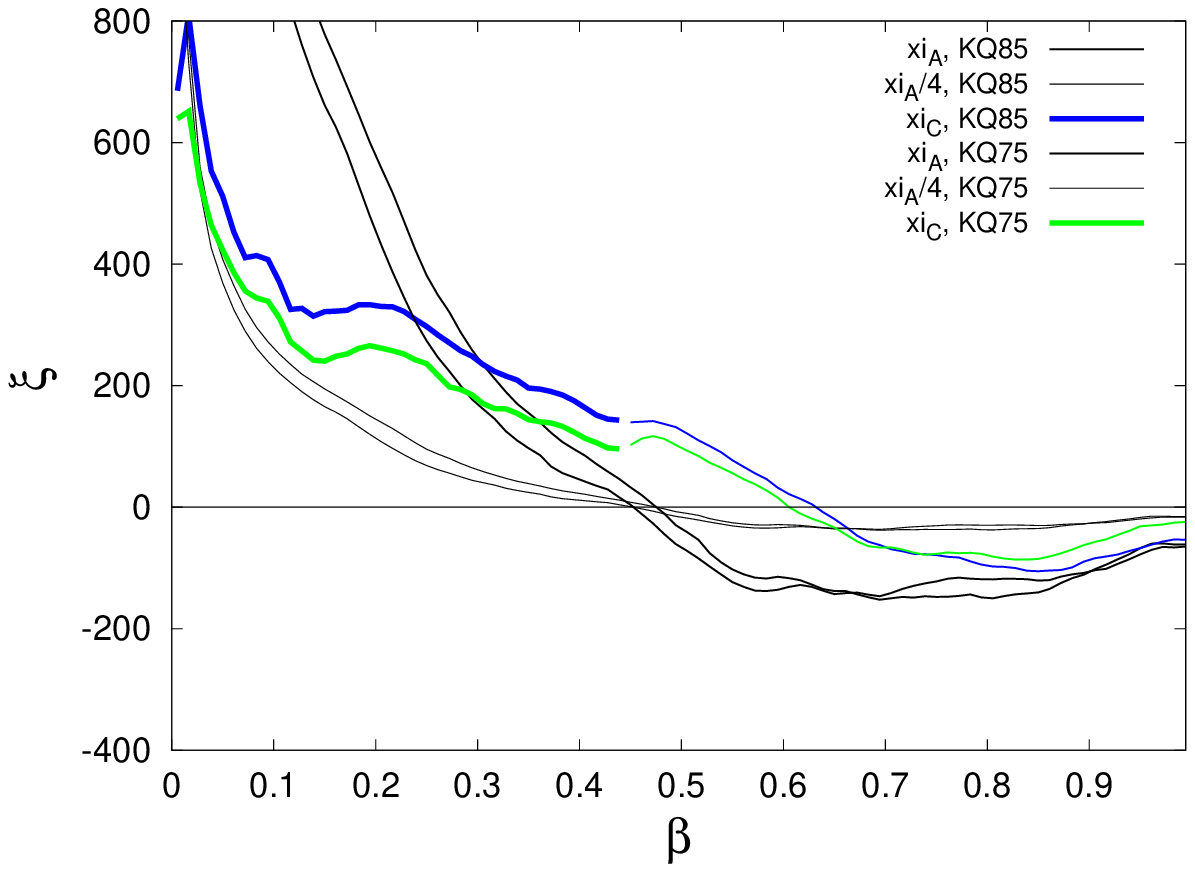}
\caption[]{ \mycaptionfont 
Spatial auto- and cross-correlation functions, $\xi_{\mathrm A}$ and $\xi_{\mathrm C}$,
respectively,
for the WMAP7 ILC map
for matched-circle radius $\alpha=23^\circ$, shown as 
{thin and
thick (colour online)}
curves, respectively. The cross-correlation function is uniformly sampled only 
for $\beta \le 0.44$ (for calculation speed); at higher separations, it is, 
in principle, biased. 
\prerefereechanges{In order to better compare the small-scale behaviour of 
$\xi_{\mathrm A}$ and $\xi_{\mathrm C}$,
the auto-correlation reduced (arbitrarily)} by a factor of 4 is also shown.
The mostly upper (lower) member of each close pair of functions is for the KQ85 (KQ75) galactic mask.
}
\label{f-corr_eg}
\end{figure} 
} 

\newcommand\fmatcheddiscs{
\begin{figure}
\centering 
\includegraphics[width=70mm]{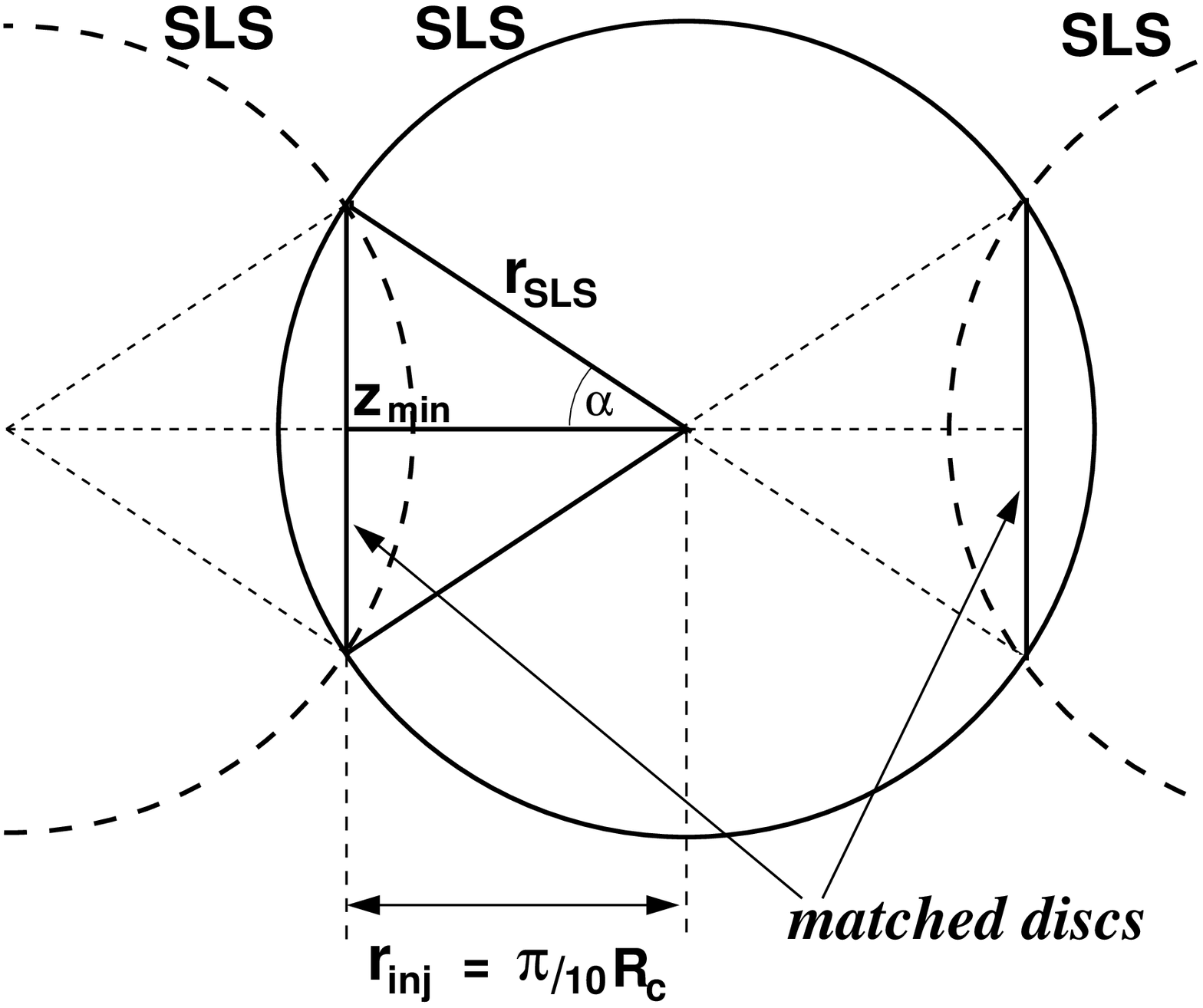} 
\caption[]{ \mycaptionfont { 
Relation of matched discs to matched circles for the Poincar\'e
dodecahedral space, with injectivity radius $\rinj = (\pi/10) \rC$,
shown in the universal covering space S$^3$ of radius $\rC$. 
Multiple copies of the surface of last scattering (SLS) of radius $\rSLS$ 
intersect in circles lying in (flat) 2-planes that are orthogonal to
the plane of the page. The interiors of the circles within these
2-planes constitute matched discs. The matched-circle (observer-centred) angular 
radius is $\alpha$.
Redshifts on either matched disc increase from $z_{\min}$ at the centre 
to $\zSLS$ on the boundary (the matched circle).
The SLS can be replaced by the 2-sphere defined by $z=200$ to obtain
matched discs with $200>z>z_{\min}$ instead of 
$\zSLS>z>z_{\min}$.
}}
\label{f-matched-discs}
\end{figure} 
} 


\section{Introduction}  \label{s-intro}

The near lack of 10~{\hGpc} scale structure in 
cosmic microwave background (CMB) temperature fluctuation maps 
was tentatively detected by 
the COsmic Background Explorer (COBE) 
and supported by the Wilkinson Microwave Anisotropy Probe (WMAP)
\nocite{WMAPbasic}({Bennett} {et~al.} \protect\hyperlink{hypertarget:WMAPbasic}{2003}). This has long been considered as a sign
that the Universe,
phenomenologically well-approximated as a Friedmann-Lema\^{\i}tre-Robertson-Walker (FLRW)
model, is spatially small. Structures bigger than 
comoving space itself cannot exist physically, and their observational statistical
description in the universal covering space (apparent space) 
should approximately reveal this \nocite{Star93,Stevens93}({Starobinsky} \protect\hyperlink{hypertarget:Star93}{1993}; {Stevens} {et~al.} \protect\hyperlink{hypertarget:Stevens93}{1993}), even though
the apparent nature of the covering space does permit structures that
are larger than the Universe (Fig.~1, \nocite{Rouk96}{Roukema} \protect\hyperlink{hypertarget:Rouk96}{1996}).
Small FLRW models have long been discussed 
\nocite{deSitt17,Fried23,Fried24,Lemaitre31ell,Rob35}(e.g.,  {de Sitter} \protect\hyperlink{hypertarget:deSitt17}{1917}; {Friedmann} \protect\hyperlink{hypertarget:Fried23}{1923}, \protect\hyperlink{hypertarget:Fried24}{1924}; {Lema{\^i}tre} \protect\hyperlink{hypertarget:Lemaitre31ell}{1927}; {Robertson} \protect\hyperlink{hypertarget:Rob35}{1935}).\footnote{\protect\postrefereechanges{\protect\nocite{Lemaitre31ell}{Lema{\^i}tre} (\protect\hyperlink{hypertarget:Lemaitre31ell}{1927}) 
is the paper in which Lema\^{\i}tre 
used redshift and distance estimates of
42 galaxies to estimate 
the value of what is now known as $H_0$, finding
575~km/s/Mpc (unweighted) or 625~km/s/Mpc (weighted).
An 
incomplete English translation that excludes almost all of 
the observational analysis 
was published} 
\protect\postrefereechanges{four years later} 
\protect\postrefereechanges{\protect\nocite{Lemaitre31elltrans}({Lema{\^i}tre} \protect\hyperlink{hypertarget:Lemaitre31elltrans}{1931}). See \protect\nocite{vdB11Lemaitre}{van den Bergh} (\protect\hyperlink{hypertarget:vdB11Lemaitre}{2011})}
\protect\postrefereechanges{and 
\protect\nocite{Block11Lemaitre}{Block} (\protect\hyperlink{hypertarget:Block11Lemaitre}{2011}) and references therein for discussion of this issue.}}
The lack of structure on the largest scales is most simply seen in the
spatial or angular two-point autocorrelation function of the temperature fluctuations,
as spherical harmonical analysis requires a ``conspiracy'' between 
different harmonics in order to match the observed weak correlation
at the largest scales \nocite{WMAPSpergel,Copi07,Copi09,Copi10,Copi10b}({Spergel} {et~al.} \protect\hyperlink{hypertarget:WMAPSpergel}{2003}; {Copi} {et~al.} \protect\hyperlink{hypertarget:Copi07}{2007}, \protect\hyperlink{hypertarget:Copi09}{2009}; {Sarkar} {et~al.} \protect\hyperlink{hypertarget:Copi10}{2011}; {Copi} {et~al.} \protect\hyperlink{hypertarget:Copi10b}{2010}). 

The locally homogeneous 3-manifolds that have 
\postrefereechanges{most frequently been studied recently}
as candidates 
of comoving space in order to fit the WMAP data with an FLRW model
include the 3-torus $T^3$
\postrefereechanges{\protect\nocite{WMAPSpergel,Aurich07align,Aurich08a,Aurich08b,Aurich09a,AslanMan11}({Spergel} {et~al.} \protect\hyperlink{hypertarget:WMAPSpergel}{2003}; {Aurich} {et~al.} \protect\hyperlink{hypertarget:Aurich07align}{2007}; {Aurich} \protect\hyperlink{hypertarget:Aurich08a}{2008}; {Aurich} {et~al.} \protect\hyperlink{hypertarget:Aurich08b}{2008}, \protect\hyperlink{hypertarget:Aurich09a}{2010}; {Aslanyan} \& {Manohar} \protect\hyperlink{hypertarget:AslanMan11}{2011})}
and the
Poincar\'e dodecahedral space $S^3/I^*$
\nocite{LumNat03,Aurich2005a,Aurich2005b,Gundermann2005,Caillerie07,RBSG08,RBG08}({Luminet} {et~al.} \protect\hyperlink{hypertarget:LumNat03}{2003}; {Aurich} {et~al.} \protect\hyperlink{hypertarget:Aurich2005a}{2005a}, \protect\hyperlink{hypertarget:Aurich2005b}{2005b}; {Gundermann} \protect\hyperlink{hypertarget:Gundermann2005}{2005}; {Caillerie} {et~al.} \protect\hyperlink{hypertarget:Caillerie07}{2007}; {Roukema} {et~al.} \protect\hyperlink{hypertarget:RBSG08}{2008b}, \protect\hyperlink{hypertarget:RBG08}{2008a}).
\postrefereechanges{A full pre-WMAP classification of the positively curved spaces
(including $S^3/I^*$)
and mathematical tools for studying them in the cosmological context were
presented by \nocite{GausSph01}{Gausmann} {et~al.} (\protect\hyperlink{hypertarget:GausSph01}{2001}).}
Some analyses suggest that the Poincar\'e space may not provide a better
fit than the infinite flat model or a hypersphere model\footnote{The hypersphere is  
frequently referred to using the ambiguous term ``closed''.}
\nocite{KeyCSS06,NJ07}({Key} {et~al.} \protect\hyperlink{hypertarget:KeyCSS06}{2007}; {Niarchou} \& {Jaffe} \protect\hyperlink{hypertarget:NJ07}{2007}). 
A heuristic argument in favour of the Poincar\'e space is that the residual 
gravity effect implied by imperfect cancellation of the gravitational signal
from distant images of a nearby massive object singles out the Poincar\'e space
as being better balanced than other spaces \nocite{RBBSJ06,RR09}({Roukema} {et~al.} \protect\hyperlink{hypertarget:RBBSJ06}{2007}; {Roukema} \& {R\'o\.za\'nski} \protect\hyperlink{hypertarget:RR09}{2009}).

If the Poincar\'e space hypothesis is correct, then the set of density
perturbations in the Universe would not constitute a statistical
ensemble of realisations of a model of density perturbation
statistics, it would consist of a {\em single physical realisation} within a
fundamental domain with a definite orientation and comoving size.
Thus, it should be possible to find the optimal orientation and size
of the fundamental domain in an astronomical coordinate system, rather
than only estimate probabilities based on ensemble averages of
simulated temperature fluctuations. 
{The parameters of the optimal solution}
should then be
usable to design observing strategies that should strengthen evidence
either for or against the hypothesis.  

\prerefereechanges{Estimates of the optimal
Poincar\'e space coordinates {have been made}
\nocite{RBSG08,RBG08}({Roukema} {et~al.} \protect\hyperlink{hypertarget:RBSG08}{2008b}, \protect\hyperlink{hypertarget:RBG08}{2008a}).}
However, 
{these estimates favoured}
the use of the moderately
large-scale signal, up to $\sim 4$~{\hGpc}. For geometrical reasons,
this had the consequence that one of the fundamental domain
parameters, the matched-circle observer-centred angular radius
$\alpha$, or equivalently, the (comoving) size of the Universe
($2\rinj$, i.e.  twice the injectivity radius, i.e. the length of the
shortest closed spatial geodesic), was, in principle, only
weakly constrained \nocite{RBSG08}(Sect.~4.2, Fig.~9, Eq.~(32),
 {Roukema} {et~al.} \protect\hyperlink{hypertarget:RBSG08}{2008b}). The random error in the estimate $\alpha \sim 21 \pm 1
\ddeg$ \nocite{RBSG08}(Table~2,  {Roukema} {et~al.} \protect\hyperlink{hypertarget:RBSG08}{2008b}) 
{was suspected to}
be dominated by a systematic
error of $\sim 10\ddeg$.

\fmatcheddiscs

This uncertainty has critical consequences for the falsifiability of
the model. 
\prerefereechanges{If 
$32^\circ \ltapprox \alpha \ltapprox 48^\circ$, then 
multiply imaged objects should be observable at 
$40 \gtapprox z \gtapprox 10$, respectively, near the centres
of {\em matched discs} (Fig.~\ref{f-matched-discs}).
Matched discs and this redshift calculation are explained
in \SSS\ref{s-matched-discs} and the results are shown in Fig.~\ref{f-zminalpha}.
This redshift range is approximately that 
when 0.1\% or so of baryons are expected to have formed H$_2$-cooled
protogalaxies within dark matter haloes of about $10^5$ to $10^6
M_{\odot}$, in which very high mass Population III stars form and some
explode as supernovae, while others may collapse directly into black
holes \nocite{Glover2005zgal30to40,CiardiF2005zgal,Greif2008zgal20}(e.g.,  {Glover} \protect\hyperlink{hypertarget:Glover2005zgal30to40}{2005}; {Ciardi} \& {Ferrara} \protect\hyperlink{hypertarget:CiardiF2005zgal}{2005}; {Greif} {et~al.} \protect\hyperlink{hypertarget:Greif2008zgal20}{2008}, and references
  therein).
Although the highest-redshift, spectroscopically confirmed, collapsed
astrophysical object is a galaxy
at $z=8.6$ \nocite{Lehnert8p6z2010}({Lehnert} {et~al.} \protect\hyperlink{hypertarget:Lehnert8p6z2010}{2010}), it is generally expected that
Population III stars may be detected as GRBs at $z > 10$ during the
coming decades.

However, if the Universe is larger, i.e. if 
$18^\circ \ltapprox \alpha \ltapprox 32^\circ$, then
the corresponding redshifts are higher, in the range
$200 \gtapprox z \gtapprox 40$, respectively, so protogalaxies can have
collapsed only in higher overdensity peaks, i.e. much more rarely. Their
rareness at higher redshifts decreases the chance that observable
objects will lie near the centres of the matched discs compared to
the case for larger values of $\alpha$.

The advantage of seeing
two topologically lensed images of an object at
nearly the same epoch is that evolutionary effects would be minimal,
although there would still be a projection effect induced by seeing the same object 
``from front and back''. Significant evidence of the absence of a predicted
{topologically lensed} 
image would be able to refute the Poincar\'e dodecahedral
space model. 
{Alternatively,} well-matched 
{topologically lensed} pairs would increase the
over-determined nature of the model, leading to successively more
and more topological lensing predictions.
Thus, it is important to see if the estimate of $\alpha$ can be made
more accurate. 
}

\fzminalpha

\prerefereechanges{A more accurate estimate of $\alpha$ 
is made by using the cross-correlation method 
\nocite{RBSG08,RBG08}({Roukema} {et~al.} \protect\hyperlink{hypertarget:RBSG08}{2008b}, \protect\hyperlink{hypertarget:RBG08}{2008a}) at higher comoving spatial resolution 
($\ltapprox 1$~{\hGpc}) than in the earlier work and updating 
to the WMAP 7-year data release. 
The method is presented in 
Sects~\ref{s-method-cross} and \ref{s-method-optimise}.}
Results are presented in \SSS\ref{s-results} and
conclusions are given in \SSS\ref{s-conclu}.
All distances are 
{FLRW}
comoving distances.
The Hubble constant is written
$H_0 = 100 h$~km/s/Mpc.

\section{Method} \label{s-method}

\prerefereechanges{
\subsection{Matched discs} \label{s-matched-discs}
}

\prerefereechanges{Here we
introduce an observational corollary of
the identified circles principle
(\nocite{Corn96,Corn98b}{Cornish} {et~al.} \protect\hyperlink{hypertarget:Corn96}{1996}, \protect\hyperlink{hypertarget:Corn98b}{1998}; see also Fig.~2, \nocite{Rouk00BASI}{Roukema} \protect\hyperlink{hypertarget:Rouk00BASI}{2000}).
Two circles on the surface of last scattering (SLS) are identified because
in the comoving covering 3-space,
they consist of the intersection of two copies of a single physical
(flat) 2-plane with a 2-sphere (the SLS). 
{Filled polygonal subsets} 
of the two copies of the 2-plane constitute {two} faces 
of a copy of the fundamental domain that can be thought of as being
pasted together. {\em The flat interiors of 
{the two matching} circles---within
the {two copies of the} 
flat 2-plane---constitute matched discs with sub-SLS
redshifts, i.e. $z_{\min} \le z \le z_{\mathrm SLS}$, where
$z_{\min}$ occurs at the disc centre.}
This is illustrated in Fig.~\ref{f-matched-discs}.

A large enough $\alpha$ (small enough $2\rinj$) would imply a low
enough $z_{\min}$ for collapsed astrophysical objects [e.g. quasars,
  Lyman-break objects or gamma-ray bursts (GRBs) from early forming
  galaxies] located near the centres of these matched discs to be seen
as nearly antipodal pairs at nearly identical redshifts. 
Objects successively further from the centres of the matched
discs would be multiply imaged at successively higher redshifts $z > z_{\min}$,
i.e. at successively earlier, but still approximately equal, epochs.
A survey whose selection criteria select a redshift band beyond $z_{\min}$
should detect an annulus of multiply imaged objects. For example,
if $z_{\min}= 20$ and the survey selects objects in the range $ 30 < z < 40$, 
then there should be an annulus of multiply imaged objects.

The 
redshift to the centres of a pair of matched discs for a given matched
circle radius $\alpha$ can be calculated by finding the value of the
dark energy parameter $\Omega_\Lambda$ that 
equates the curvature radius calculated locally with its value
calculated globally using the injectivity radius.
The former is $\rC = (c/H_0) \left(\Omm + \Omega_\Lambda
-1\right)^{-0.5}$, where $c$ is the spacetime unit conversion
constant, $\Omm$ is a fixed value of the matter density parameter, and $H_0$ is the
Hubble constant. 
The calculation using the injectivity radius is made using Eq.~(15) of
\nocite{RBSG08}{Roukema} {et~al.} (\protect\hyperlink{hypertarget:RBSG08}{2008b}), $\alpha$ and the comoving distance to the surface
of last scattering $\rSLS (\Omm,\Omega_\Lambda,H_0,\zSLS)$, assuming
$z_{\mathrm{SLS}} = 1100$. Thus,
\begin{equation}
\rC = \frac{c}{H_0} \frac{1}{\sqrt{\Omega_\Lambda + \Omm - 1}} 
= \frac{\rSLS(\Omm,\Omega_\Lambda,H_0)}{
  \arctan \left[ \frac{\tan (\pi/10)}{\cos \alpha} \right] }.
\label{e-rC-equality}
\end{equation}
Given $\Omm$ and $H_0$, the latter equality 
can be solved numerically for $\Omega_\Lambda$, yielding $\rC$
from the former equality.
The injectivity radius is 
\begin{equation}
\rinj= (\pi/10)\rC,
\label{e-rinj-poincare}
\end{equation}
so the redshift $z_{\min}$ to 
the centre of the matched discs is obtained by inverting 
\begin{equation}
\rinj = r(\Omm,\Omega_\Lambda,H_0,z_{\min})
\label{e-standardFLRW-rradial}
\end{equation}
where $r$ is the standard FLRW radial comoving distance.\footnote{It might
be useful to call $z_{\min}$ the injectivity redshift 
$z_{\mathrm{inj}}$.}

Figure~\ref{f-zminalpha} (upper panel) shows that if $32^\circ
\ltapprox \alpha \ltapprox 48^\circ$, then the redshifts of objects
lying in the matched discs would be in the range $40 \gtapprox z
\gtapprox 10$, respectively.
For smaller matched circle radii 
$18^\circ \ltapprox \alpha \ltapprox 32^\circ$, 
the corresponding redshifts are higher, in the range
$200 \gtapprox z \gtapprox 40$, respectively.

\postrefereechanges{Larger matched circle radii correspond to lower
redshifts, but also higher total density and smaller curvature radii
$\rC$. For example, Fig.~15 of \nocite{GausSph01}{Gausmann} {et~al.} (\protect\hyperlink{hypertarget:GausSph01}{2001}) 
shows a multiple imaging spike with a minimum redshift
lying in the range $1 < z < 3$ when
$\Omm=0.35, \Omega_\Lambda=0.75$. This can be calculated by using the
left-hand equality of Eq.~(\ref{e-rC-equality}) to obtain $\rC = 9.48{\hGpc}$, 
Eq.~(\ref{e-rinj-poincare}) to obtain $\rinj = 2.98{\hGpc}$, and inverting 
Eq.~(\ref{e-standardFLRW-rradial}), yielding $z=1.45$. 
Use of 
the right-hand expression in Eq.~(\ref{e-rC-equality}) shows that
the corresponding matched circle radius is $\alpha = 77^\circ$.}

The lower panel of Fig.~\ref{f-zminalpha} shows the
corresponding injectivity radius $\rinj$ of the Universe. While
the redshift at the centre of a matched disc is {insensitive}
to $\Omm$, the size of the Universe is sensitive to
$\Omm$.  When $\alpha = 0^\circ$, the injectivity radius is identical
to the comoving distance to the surface of last scattering, i.e.
$\rinj = \rSLS$ [cf. \postrefereechanges{Eqs~(\ref{e-rC-equality}) and 
(\ref{e-rinj-poincare})}].
}

\subsection{Cross-correlation maximisation method} \label{s-method-cross}
To improve the estimate of $\alpha$, the cross-correlation method 
\nocite{RBSG08,RBG08}({Roukema} {et~al.} \protect\hyperlink{hypertarget:RBSG08}{2008b}, \protect\hyperlink{hypertarget:RBG08}{2008a}) is used over the single parameter $\alpha$,
using the previously found solution for the positions of the axes
of the fundamental domain. The cross-correlation method
can be thought of as reconstructing the spatial two-point correlation
function of density fluctuations (using temperature fluctuations as
a proxy) at {\em sub-}injectivity radius length scales, where an observed
pair of points is only used if the observed spatial separation in 
the comoving covering space is {\em larger} than the injectivity radius.
Holonomies $g_i$ are applied to one member $y$ of a pair of observed 
spatial points $(x,y)$ so that if the implied pair $(x,g_i^{-1}(y))$
is a close pair, then the observed temperature fluctuation product 
$\Delta T(x)\,\Delta T(y)$ 
{is cumulated} to the cross-correlation
{estimate}
at the separation $d[x,g_i^{-1}(y)]$. The method can also 
be thought of as an extension of the identified circles principle
\nocite{Corn96,Corn98b}({Cornish} {et~al.} \protect\hyperlink{hypertarget:Corn96}{1996}, \protect\hyperlink{hypertarget:Corn98b}{1998}),
but with a uniform sky weighting over observed points instead of 
a weighting that favours matched circle intersections.
If a hypothesised set of 
sky orientation, injectivity radius and twist parameters are
physically correct, then the locally isotropic components of 
the auto-correlation function 
should be reproduced by the cross-correlation function.

\subsection{Correlation functions and optimisation criterion} \label{s-method-optimise}
The Markov chain Monte Carlo (MCMC) method used in
\nocite{RBSG08,RBG08}{Roukema} {et~al.} (\protect\hyperlink{hypertarget:RBSG08}{2008b}, \protect\hyperlink{hypertarget:RBG08}{2008a}) adopted $\Npoint = 2000$ points per position in
MCMC parameter space \nocite{RBSG08}(Sect.~3.5,  {Roukema} {et~al.} \protect\hyperlink{hypertarget:RBSG08}{2008b}).  
\prerefereechanges{The optimisation
criterion (a pseudo-probability) was
aimed at maximising 
$\xi_{\mathrm C} - \xi_{\mathrm A}$ at $\ltapprox 4$~{\hGpc}
scales.  The definition of the criterion [Eqs~(25), (26), \nocite{RBSG08}{Roukema} {et~al.} \protect\hyperlink{hypertarget:RBSG08}{2008b}] 
took into account the
uncertainty in estimating the cross-correlation $\ximc(r)$ induced by
using a small value of $\Npoint$.  This provided a
practical method of exploring a large volume of parameter space,
focussing on cross- and auto-correlations mostly at the upper end of
the separation range $r \ltapprox 4$~{\hGpc}.}

\fxiilcmain

\fxiV

\fxiW

\fxiilcalt

Here, in order to optimise the estimate of $\alpha$, a much higher
number of points is used per correlation calculation, i.e.  $\Npoint =
2\times 10^{5}$, and only $\alpha$ is varied. This enables the use of
the correlation signal at separations $r \ltapprox 1$~{\hGpc}.
From Fig.~11
of \nocite{RBSG08}{Roukema} {et~al.} (\protect\hyperlink{hypertarget:RBSG08}{2008b}), it is clear that $\xisc \gg \ximc$ at 
$r \ltapprox 1$~{\hGpc}.
Because of the complicated mix of different effects---the
na\"{\i}ve Sachs-Wolfe effect, the integrated Sachs-Wolfe effect, the
Doppler effect, and intrinsic temperature fluctuations---that
contribute to $\ximc$ and the auto-correlation $\xisc$, it is not
obvious \prerefereechanges{whether 
$\ximc$ should be as high as $\xisc$ at sub-Gpc
scales. In other words,
on these smaller scales, the increasing contribution of 
the Doppler effect---which is locally anisotropic---to the temperature
fluctuations makes it less realistic to test the 
hypothesis that $\ximc$ and $\xisc$ are identical functions apart
from calculational uncertainty.
Moreover, the higher value of $\Npoint$ reduces
the numerical usefulness of estimating the calculational uncertainty.}

Hence, 
\prerefereechanges{with the aim of maximising the cross-correlation of
locally isotropic emission,}
the optimisation criterion is defined as the mean
cross-correlation below a given length 
\prerefereechanges{scale, without a direct
comparison with the auto-correlation. 
However, 
since the comoving (spatial geodesic) separation
$d(\theta_{\mathrm A},\rSLS)$ between two fixed points on
the SLS separated by an observer-centred angle $\theta_{\mathrm A}$ varies
with $\Omm$ and $\Omega_\Lambda$, $d$ needs to be replaced by a scaled
separation $\beta$ that to a good approximation depends only on $\theta_{\mathrm A}$,
in order that
$\xisc(\beta)$
has only a weak dependence on $\Omm$ and $\Omega_{\Lambda}$.
This can be done for small separations $d$, which are the ones of most interest.
As in Eq.~(16) of \nocite{RBSG08}{Roukema} {et~al.} (\protect\hyperlink{hypertarget:RBSG08}{2008b}), the
spherical sine law gives
\begin{equation}
\frac{   \sin[ d(\theta_{\mathrm A},\rSLS) /(2\rC)]}{\rSLS} 
  =   \frac{1}{\rSLS}
  \sin \frac{\theta_{\mathrm A}}{2} 
  \sin\frac{\rSLS}{\rC}
  \label{e-autocorr-separation-A}
  ,
\end{equation}
i.e.
\begin{equation}
\frac{   d(\theta_{\mathrm A},\rSLS)}{\rSLS} 
  \approx    2\frac{\rC}{\rSLS}
  \sin\frac{\rSLS}{\rC}
  \sin \frac{\theta_{\mathrm A}}{2} 
  ,
  \label{e-autocorr-separation-B}
\end{equation}
where the second and higher order terms in the Taylor expansion of 
$\sin[d/(2\rC)]$ are dropped, since for $d \ltapprox 1\,${\hGpc} 
and $\rC \gtapprox 25\,${\hGpc}, $d/(2\rC) \ltapprox 0.02$.
Using Eq.~(15) of \nocite{RBSG08}{Roukema} {et~al.} (\protect\hyperlink{hypertarget:RBSG08}{2008b}), the ratio
$\frac{\rSLS}{\rC}$ depends only on $\alpha$,
without any direct dependence on $\Omm$ and $\Omega_\Lambda$. 
For $0^\circ \le \alpha \le 45^\circ$, we have
$1.967 \ge 2\frac{\rC}{\rSLS}  \sin\frac{\rSLS}{\rC} \ge 1.939$, respectively,
i.e. the latter expression varies by less than 1.5\% over this range.
In contrast, for $0.25 \le \Omm \le 0.35$, $\rSLS$ varies by about 14\%.
Thus, $\beta := d/\rSLS$ depends mostly on $\theta_{\mathrm A}$ 
and only weakly on $\Omm$ and $\Omega_\Lambda$.

This is used for the mean
cross-correlation, i.e. the latter is calculated below a fixed
fraction $\beta$ of the comoving distance to the 
surface of last scattering radius,
\begin{equation}
  \bar\xi_{\beta}(\alpha) := \frac{1}{\beta \rSLS} 
  \int_0^{\beta \rSLS} \ximc(r,\alpha) \mathrm{d}r. 
\label{e-defn-xim}
\end{equation}
The primary scales of interest are
$\beta = 0.033$ and $0.1$, corresponding to $\beta \rSLS \approx 0.33$ and 
$1$~{\hGpc}, respectively.}

\fcorreg

The primary estimate is the matched circle radius
$\alpha$ that maximises
\prerefereechanges{$\bar\xi_{\beta}(\alpha)$}
 for $\beta = 0.033$ and $0.1$ for the
7-year WMAP ILC 
map,\footnote{\href{http://lambda.gsfc.nasa.gov/data/map/dr4/dfp/ilc/wmap_ilc_7yr_v4.fits}{{\tt http://lambda.gsfc.nasa.gov/data/map/dr4/dfp/}}
\href{http://lambda.gsfc.nasa.gov/data/map/dr4/dfp/ilc/wmap_ilc_7yr_v4.fits}{{\tt ilc/wmap\_ilc\_7yr\_v4.fits}}} masking with the KQ85 mask to minimise galactic
contamination.
Estimates of the systematic error, assumed to dominate the
random error, are made by 
considering
the foreground-reduced V band\footnote{\href{http://lambda.gsfc.nasa.gov/data/map/dr4/skymaps/7yr/forered/wmap_band_forered_iqumap_r9_7yr_V_v4.fits}{{\tt http://lambda.gsfc.nasa.gov/data/map/dr4/skymaps/}} 
\href{http://lambda.gsfc.nasa.gov/data/map/dr4/skymaps/7yr/forered/wmap_band_forered_iqumap_r9_7yr_V_v4.fits}{{\tt 7yr/forered/wmap\_band\_forered\_iqumap\_r9\_7yr\_V\_v4.fits}}} 
and 
 W band maps,\footnote{\href{http://lambda.gsfc.nasa.gov/data/map/dr4/skymaps/7yr/forered/wmap_band_forered_iqumap_r9_7yr_W_v4.fits}{{\tt http://lambda.gsfc.nasa.gov/data/map/dr4/skymaps/}} 
\href{http://lambda.gsfc.nasa.gov/data/map/dr4/skymaps/7yr/forered/wmap_band_forered_iqumap_r9_7yr_W_v4.fits}{{\tt 7yr/forered/wmap\_band\_forered\_iqumap\_r9\_7yr\_W\_v4.fits}}}
and by applying the KQ75 mask.

\talphabest

Since the aim here is to 
{see whether the uncertainty in $\alpha$ can
be reduced so that it no longer dominates the uncertainty in the 
optimal solution}, 
the sky orientation parameters of the fundamental domain are fixed
{at $(l,b,\theta) = (184.0^\circ, 62.0^\circ, 34.0^\circ)$,} 
using the exact dodecahedral solution found in 
\nocite{RBSG08}{Roukema} {et~al.} (\protect\hyperlink{hypertarget:RBSG08}{2008b}) (Fig.~11 caption) rather than the individual axis estimates.


\section{Results} \label{s-results}

\subsection{Matched circle radius $\alpha$}

Figure~\ref{f-xi-ilc-85} shows a well-defined local peak in the mean
correlation \prerefereechanges{$\bar\xi_{\beta}(\alpha)$}
near $\alpha = 23^\circ$.
Even for $\beta=0.4$, i.e. using the $\ltapprox 4$~{\hGpc}
cross-correlations, a small local peak near the same value is
present. The smaller scale cross-correlations clearly overcome
the concern of a systematic error of about $10^\circ$ in the estimate
of $\alpha$.

Several features of this
figure can be interpreted simply if the overall Poincar\'e dodecahedral
space solution is physically correct.
Firstly, as would be expected for a true solution, the rough solution found using
the larger scale signal becomes better defined using the smaller scale
signal. Secondly, as $\alpha$ decreases down to $10^\circ$, 
\prerefereechanges{$\bar\xi_{\beta}(\alpha)$}
drops strongly. The smaller $\beta$ is, 
the sharper the drop. This is consistent with the solution being correct,
since matched circles for wrong values of $\alpha$ that are smaller than
the would-be true value force small-scale cross-correlations to be estimated
at large true separations, where the auto-correlation is weak.

Figures~\ref{f-xi-V-85} and \ref{f-xi-W-85} show that the
peak in 
\prerefereechanges{$\bar\xi_{\beta}(\alpha)$} is present in both the V and W
foreground-reduced maps. Figure~\ref{f-xi-ilc-75} shows that the 
peak is a little weakened using a stronger galactic cut, but remains
present. Because of the
$36^\circ$ rotation {used in matching circles or discs}, 
stronger galactic cuts severely
reduce the numbers {of} matched pairs and nearly matched pairs.

The matched circle radii that maximise 
\prerefereechanges{$\bar\xi_{\beta}$},
{after a Gaussian smoothing with $\sigma_\alpha=0.2^\circ$,}
are
listed in Table~\ref{t-alphabest}. 
{In the ILC KQ85-masked map,
the maximum cross-correlation is independent of
the integration length scale $\beta$, giving $\alpha=23.0^\circ$
in all four cases. A sharp maximum cross-correlation is clearly present 
at about the same matched circle radius for 
the single-band V and W (foreground-reduced) maps, but with some dependence
on $\beta$. The $\ltapprox 1$~{\hGpc} scale V and W maps 
($\beta=0.033, 0.1$) give $21.8^\circ \le \alpha \le 23^\circ$.
The ILC KQ75-masked map again gives a maximum 
\prerefereechanges{$\bar\xi_{\beta}(\alpha)$}
that is independent of $\beta$, i.e. $\alpha=23.8^\circ$, a little higher than the other estimates. 
Conservatively taking the maximum differences to be independent and
adding them in quadrature, the systematic error in 
the estimate of $\alpha$ induced by the choice of the map and the galactic
mask is $1.4^\circ$, giving $\alpha=23.0\pm1.4^\circ$ as the 
improved estimate of the matched circle radius.}

The cross-correlation functions for the ILC map for 
{$\alpha=23^\circ$} are
shown in Fig.~\ref{f-corr_eg}.  This should be compared with Fig.~11 
of \nocite{RBSG08}{Roukema} {et~al.} (\protect\hyperlink{hypertarget:RBSG08}{2008b}). In Fig.~\ref{f-corr_eg}, 
since the method of \nocite{RBG08}{Roukema} {et~al.} (\protect\hyperlink{hypertarget:RBG08}{2008a}) is used to speed
up the calculations, only separations below $\beta=0.44$, i.e. about 
4.4~{\hGpc}, are based on an underlying uniform sky distribution.
Nevertheless, cross-correlations at higher scales are similar to those
in Fig.~11 of \nocite{RBSG08}{Roukema} {et~al.} (\protect\hyperlink{hypertarget:RBSG08}{2008b}), i.e. they do not appear to
be strongly affected by the potential bias.

The main new feature in Fig.~\ref{f-corr_eg} is that there is a
sharp upturn at 
{sub-Gpc scales, 
which was not visible
in Fig.~11 of \nocite{RBSG08}{Roukema} {et~al.} (\protect\hyperlink{hypertarget:RBSG08}{2008b}),
rather than a continuation
of the approximately constant slope on 1--5~{\hGpc} scales, visible
in both figures.}
\prerefereechanges{If the na\"{\i}ve Sachs-Wolfe effect, i.e. a  
locally isotropic effect, contributes about
half of typical temperature fluctuation amplitudes
on these scales, then the expected cross-correlation should be 
$\xi_{\mathrm C} \sim \xi_{\mathrm A}/4$.
The latter function is plotted 
in Fig.~\ref{f-corr_eg}, in addition to $\xi_{\mathrm A}$.
This shows that 
{$\xi_{\mathrm C} \sim \xi_{\mathrm A}/4$} indeed gives a rough
estimate of the cross-correlation on sub-Gpc scales
{with $\beta > 0.011$, 
i.e., $\beta \rSLS \gtapprox 0.11$~{\hGpc}.}
Thus, a possible order-of-magnitude interpretation would be that about half the 
temperature fluctuations on these scales are locally isotropic.}

\subsection{Inferred parameters: $2\rinj, z_{\mathrm{min}}, \alpha_{200}$}

{As 
described above, for a choice of $\Omm$, $H_0,$ and $\alpha$, 
Eqs~(\ref{e-rC-equality}) and (\ref{e-rinj-poincare})
can be used to find $\Omega_\Lambda$, $\rC$,
$\rinj$ and the redshift to the matched disc centres $z_{\mathrm{min}}$.
For $\Omm=0.28\pm0.02$ and $\alpha=23\pm1.4 ^\circ$, and assuming that these give
Gaussian independent errors, twice the injectivity radius of the Universe
is $2\rinj = 18.2\pm0.5$~{\hGpc}, and the redshift to the matched disc
centres is $z_{\mathrm{min}} = 106\pm18$.}

The angular radius of a matched disc from its centre
at $z_{\mathrm{min}} = 106\pm18$ up to a redshift at which some of the
very high density peaks should first collapse, $z=200$, can be calculated
analogously to Eq.~(15) of \nocite{RBSG08}{Roukema} {et~al.} (\protect\hyperlink{hypertarget:RBSG08}{2008b}), i.e. 
\begin{equation}
  \alpha_{200} = \arccos \left[
  \cos\alpha \, \frac{\tan (\rSLS/\rC)}{\tan (r_{200}/\rC)} \right],
  \label{e-alpha-200}
\end{equation}
where $r_{200}$ is the comoving radial distance to $z=200$ for the
same values of $\Omm,\Omega_\Lambda,$ and $H_0$ as for the other parameters.
This yields 
{$\alpha_{200} = 14.8\pm2.3^\circ$.} 
{These discs} do not intersect with one another. Thus, 
for the full sky, the 12 matched discs for $200 > z > 106$ would
project to a fraction of 
{$24 \pi(1 - \cos\alpha_{200} )/(4\pi)$ = $20\pm6\%$}
of the sky. The fractional coverage of the unmasked sky should be similar.

\section{Conclusion} \label{s-conclu}

The sub-Gpc cross-correlation estimates 
shown in Figs~\ref{f-xi-ilc-85}--\ref{f-xi-ilc-75} 
show that the systematic uncertainty in the earlier estimate of the
matched circle radius for the optimal fit of the Poincar\'e dodecahedral
space model to WMAP sky maps is significantly reduced 
by using this high-resolution signal, giving
{$\alpha = 23 \pm 1.4\ddeg$.}
For $\Omm=0.28\pm0.02$, the inferred size of the Universe
is $2\rinj = 18.2\pm0.5$~{\hGpc}. 

This constraint on $\alpha$ pushes {sub-SLS} multiple-image testing of
the Poincar\'e space model (at least) several decades into the future,
given present plans for new telescope projects. Nevertheless, higher
density peaks should collapse and form population III stars earlier
than lower density peaks, so future surveys of very high density peaks
that collapse early, during $200> z > 106$, should, in principle, be
feasible.  The objects would be (multiply) visible in 12 matched flat
discs that project to about 20\% of the full sky. The redshifts of the
objects increase radially from 
{$z=106\pm18$} at the centre of a disc 
\prerefereechanges{out to the edge of the disc defined by
$z=200$. An annulus extending the disc further to $\zSLS \approx 1100$
on the SLS would define a multiply imaged region where collapsed
objects would be extremely rare.}
The two
topologically lensed images {of a} physical object are located
at identical redshifts at widely \prerefereechanges{separated}
directions in the sky in discs that are \prerefereechanges{antipodal} and
matched with a twist of $+36^\circ$ (right-handed). The redshift
equality of the two images of an object implies that evolutionary
effects would not be a problem for objects located exactly in the
discs. Objects that are in the close foreground/background (foreground
in one disc, background in the other disc) would have small
evolutionary differences between the two images of an object.

\begin{acknowledgements}
A part of this project has made use of 
Program Oblicze\'n Wielkich Wyzwa\'n nauki i techniki (POWIEW)
computational resources (grant 87) at the Pozna\'n 
Supercomputing and Networking Center (PCSS).
Use was made of the Centre de Donn\'ees astronomiques de Strasbourg 
(\url{http://cdsads.u-strasbg.fr}),
the GNU {\sc plotutils} graphics package,
and
the GNU {\sc Octave} command-line, high-level numerical computation software 
(\url{http://www.gnu.org/software/octave}). 

%
%

\end{acknowledgements}

\subm{ \clearpage }

\nice{
%

}




\begin{thebibliography}{}

\bibitem[{Aslanyan} \& {Manohar} (2011)]{AslanMan11}
\hypertarget{hypertarget:AslanMan11}{{Aslanyan},} G., \& {Manohar}, A.~V. 2011, ArXiv e-prints, \eprint{1104.0015}

\bibitem[{Aurich} (2008)]{Aurich08a}
\hypertarget{hypertarget:Aurich08a}{{Aurich},} R. 2008, \cqg, 25, 225017, \eprint{0803.2130}

\bibitem[{Aurich}, {Janzer}, {Lustig}, \&  {Steiner} (2008)]{Aurich08b}
\hypertarget{hypertarget:Aurich08b}{{Aurich}, R., {Janzer}, H.~S., {Lustig}, S., \& {Steiner},} F. 2008, Classical  and Quantum Gravity, 25, 125006, \eprint{0708.1420}

\bibitem[{Aurich}, {Lustig}, \&  {Steiner} (2005a)]{Aurich2005a}
\hypertarget{hypertarget:Aurich2005a}{{Aurich}, R., {Lustig}, S., \& {Steiner}, F. 2005a, \cqg,} 22,  3443, \eprint{astro-ph/0504656}

\bibitem[{Aurich}, {Lustig}, \&  {Steiner} (2005b)]{Aurich2005b}
\hypertarget{hypertarget:Aurich2005b}{{Aurich}, R., {Lustig}, S., \& {Steiner}, F. 2005b, \cqg,} 22,  2061, \eprint{astro-ph/0412569}

\bibitem[{Aurich}, {Lustig}, \& {Steiner} (2010)]{Aurich09a}
\hypertarget{hypertarget:Aurich09a}{{Aurich}, R., {Lustig}, S., \& {Steiner}, F. 2010, \cqg,} 27, 095009,  \eprint{0903.3133}

\bibitem[{Aurich}, {Lustig}, {Steiner}, \&  {Then} (2007)]{Aurich07align}
\hypertarget{hypertarget:Aurich07align}{{Aurich}, R., {Lustig}, S., {Steiner}, F., \& {Then},} H. 2007, \cqg, 24, 1879,  \eprint{astro-ph/0612308}

\bibitem[{Bennett}, {Halpern}, {Hinshaw}, {Jarosik},  {Kogut}, {Limon}, {Meyer}, {Page}, {Spergel}, {Tucker}, {Wollack}, {Wright},  {Barnes}, {Greason}, {Hill}, {Komatsu}, {Nolta}, {Odegard}, {Peiris},  {Verde}, \& {Weiland} (2003)]{WMAPbasic}
\hypertarget{hypertarget:WMAPbasic}{{Bennett}, C.~L., {Halpern}, M., {Hinshaw}, G., {et al.} 2003,} \apjs, 148, 1,  \eprint{astro-ph/0302207}

\bibitem[{Block} (2011)]{Block11Lemaitre}
\hypertarget{hypertarget:Block11Lemaitre}{{Block}, D.~L. 2011, ArXiv e-prints,} \eprint{1106.3928}

\bibitem[{Caillerie}, {Lachi{\`e}ze-Rey}, {Luminet},  {Lehoucq}, {Riazuelo}, \& {Weeks} (2007)]{Caillerie07}
\hypertarget{hypertarget:Caillerie07}{{Caillerie}, S., {Lachi{\`e}ze-Rey}, M., {Luminet}, J.-P., {et al.} 2007,} \aap, 476, 691, \eprint{0705.0217v2}

\bibitem[{Ciardi} \& {Ferrara} (2005)]{CiardiF2005zgal}
\hypertarget{hypertarget:CiardiF2005zgal}{{Ciardi}, B., \& {Ferrara}, A. 2005, \ssr, 116, 625,} \eprint{astro-ph/0409018}

\bibitem[{Copi}, {Huterer}, {Schwarz}, \&  {Starkman} (2007)]{Copi07}
\hypertarget{hypertarget:Copi07}{{Copi}, C.~J., {Huterer}, D., {Schwarz}, D.~J., \& {Starkman},} G.~D. 2007,  \prd, 75, 023507, \eprint{astro-ph/0605135}

\bibitem[{Copi}, {Huterer}, {Schwarz}, \&  {Starkman} (2009)]{Copi09}
\hypertarget{hypertarget:Copi09}{{Copi}, C.~J., {Huterer}, D., {Schwarz}, D.~J., \& {Starkman},} G.~D. 2009,  \mnras, 399, 295, \eprint{0808.3767}

\bibitem[{Copi}, {Huterer}, {Schwarz}, \&  {Starkman} (2010)]{Copi10b}
\hypertarget{hypertarget:Copi10b}{{Copi}, C.~J., {Huterer}, D., {Schwarz}, D.~J., \& {Starkman},} G.~D. 2010,  Advances in Astronomy, 2010, 847541, \eprint{1004.5602}

\bibitem[{Cornish}, {Spergel}, \& {Starkman} (1996)]{Corn96}
\hypertarget{hypertarget:Corn96}{{Cornish}, N.~J., {Spergel}, D.~N., \& {Starkman}, G.~D. 1996, ArXiv Gen.Rel.  \& Quant.Cosm. e-prints,} \eprint{gr-qc/9602039}

\bibitem[{Cornish}, {Spergel}, \&  {Starkman} (1998)]{Corn98b}
\hypertarget{hypertarget:Corn98b}{{Cornish}, N.~J., {Spergel}, D.~N., \& {Starkman}, G.~D. 1998, \cqg,} 15, 2657

\bibitem[{de Sitter} (1917)]{deSitt17}
\hypertarget{hypertarget:deSitt17}{{de Sitter}, W. 1917,} \mnras, 78, 3

\bibitem[{Friedmann} (1923)]{Fried23}
\hypertarget{hypertarget:Fried23}{{Friedmann},} A. 1923, {{\sl Mir kak prostranstvo i vremya} (The Universe as  Space and Time)} (Leningrad: Academia)

\bibitem[{Friedmann} (1924)]{Fried24}
\hypertarget{hypertarget:Fried24}{{Friedmann}, A. 1924,} Zeitschr. f\"ur Phys., 21, 326

\bibitem[{Gausmann}, {Lehoucq}, {Luminet}, {Uzan}, \&  {Weeks} (2001)]{GausSph01}
\hypertarget{hypertarget:GausSph01}{{Gausmann}, E., {Lehoucq}, R., {Luminet}, J.-P., {Uzan},} J.-P., \& {Weeks}, J.  2001, \cqg, 18, 5155, \eprint{gr-qc/0106033}

\bibitem[{Glover} (2005)]{Glover2005zgal30to40}
\hypertarget{hypertarget:Glover2005zgal30to40}{{Glover},} S. 2005, \ssr, 117, 445, \eprint{astro-ph/0409737}

\bibitem[{Greif}, {Johnson}, {Klessen}, \&  {Bromm} (2008)]{Greif2008zgal20}
\hypertarget{hypertarget:Greif2008zgal20}{{Greif}, T.~H., {Johnson}, J.~L., {Klessen}, R.~S., \& {Bromm},} V. 2008,  \mnras, 387, 1021, \eprint{0803.2237}

\bibitem[{Gundermann} (2005)]{Gundermann2005}
\hypertarget{hypertarget:Gundermann2005}{{Gundermann}, J. 2005, ArXiv e-prints,} \eprint{astro-ph/0503014}

\bibitem[{Key}, {Cornish}, {Spergel}, \&  {Starkman} (2007)]{KeyCSS06}
\hypertarget{hypertarget:KeyCSS06}{{Key}, J.~S., {Cornish}, N.~J., {Spergel}, D.~N., \& {Starkman},} G.~D. 2007,  \prd, 75, 084034, \eprint{astro-ph/0604616}

\bibitem[{Lehnert}, {Nesvadba}, {Cuby}, {Swinbank},  {Morris}, {Cl{\'e}ment}, {Evans}, {Bremer}, \& {Basa} (2010)]{Lehnert8p6z2010}
\hypertarget{hypertarget:Lehnert8p6z2010}{{Lehnert}, M.~D., {Nesvadba}, N.~P.~H., {Cuby}, J.-G., {et al.} 2010,} \nat, 467, 940, \eprint{1010.4312}

\bibitem[{Lema{\^i}tre} (1927)]{Lemaitre31ell}
\hypertarget{hypertarget:Lemaitre31ell}{{Lema{\^i}tre}, G. 1927,} Annales de la Soci\'et\'e Scientifique de Bruxelles,  47, 49

\bibitem[{Lema{\^i}tre} (1931)]{Lemaitre31elltrans}
\hypertarget{hypertarget:Lemaitre31elltrans}{{Lema{\^i}tre}, G. 1931,} \mnras, 91, 483

\bibitem[{Luminet}, {Weeks}, {Riazuelo}, {Lehoucq}, \&  {Uzan} (2003)]{LumNat03}
\hypertarget{hypertarget:LumNat03}{{Luminet}, J., {Weeks}, J.~R., {Riazuelo}, A., {Lehoucq},} R., \& {Uzan}, J.  2003, \nat, 425, 593, \eprint{astro-ph/0310253}

\bibitem[{Niarchou} \& {Jaffe} (2007)]{NJ07}
\hypertarget{hypertarget:NJ07}{{Niarchou}, A., \& {Jaffe}, A. 2007, \PRL, 99, 081302,}  \eprint{astro-ph/0702436}

\bibitem[{Robertson} (1935)]{Rob35}
\hypertarget{hypertarget:Rob35}{{Robertson}, H.~P. 1935,} \apj, 82, 284

\bibitem[{Roukema} (1996)]{Rouk96}
\hypertarget{hypertarget:Rouk96}{{Roukema},} B.~F. 1996, \mnras, 283, 1147, \eprint{astro-ph/9603052}

\bibitem[{Roukema} (2000)]{Rouk00BASI}
\hypertarget{hypertarget:Rouk00BASI}{{Roukema},} B.~F. 2000, \BASI, 28, 483, \eprint{astro-ph/0010185}

\bibitem[{Roukema}, {Bajtlik}, {Biesiada},  {Szaniewska}, \& {Jurkiewicz} (2007)]{RBBSJ06}
\hypertarget{hypertarget:RBBSJ06}{{Roukema}, B.~F., {Bajtlik}, S., {Biesiada}, M., {Szaniewska},} A., \&  {Jurkiewicz}, H. 2007, \aap, 463, 861, \eprint{astro-ph/0602159}

\bibitem[{Roukema}, {Buli\'nski}, \&  {Gaudin} (2008a)]{RBG08}
\hypertarget{hypertarget:RBG08}{{Roukema}, B.~F., {Buli\'nski}, Z., \& {Gaudin}, N.~E. 2008a,  \aap,} 492, 673, \eprint{0807.4260}

\bibitem[{Roukema}, {Buli\'nski},  {Szaniewska}, \& {Gaudin} (2008b)]{RBSG08}
\hypertarget{hypertarget:RBSG08}{{Roukema}, B.~F., {Buli\'nski}, Z., {Szaniewska}, A., \& {Gaudin},} N.~E.  2008b, \aap, 486, 55, \eprint{0801.0006}

\bibitem[{Roukema} \& {R\'o\.za\'nski} (2009)]{RR09}
\hypertarget{hypertarget:RR09}{{Roukema}, B.~F., \& {R\'o\.za\'nski}, P.~T. 2009, \aap, 502, 27,}  \eprint{0902.3402}

\bibitem[{Sarkar}, {Huterer}, {Copi}, {Starkman}, \&  {Schwarz} (2011)]{Copi10}
\hypertarget{hypertarget:Copi10}{{Sarkar}, D., {Huterer}, D., {Copi}, C.~J., {Starkman},} G.~D., \& {Schwarz},  D.~J. 2011, Astroparticle Physics, 34, 591, \eprint{1004.3784}

\bibitem[{Spergel}, {Verde}, {Peiris}, {Komatsu},  {Nolta}, {Bennett}, {Halpern}, {Hinshaw}, {Jarosik}, {Kogut}, {Limon},  {Meyer}, {Page}, {Tucker}, {Weiland}, {Wollack}, \& {Wright} (2003)]{WMAPSpergel}
\hypertarget{hypertarget:WMAPSpergel}{{Spergel}, D.~N., {Verde}, L., {Peiris}, H.~V., {et al.} 2003,} \apjs, 148, 175,  \eprint{astro-ph/0302209}

\bibitem[{Starobinsky} (1993)]{Star93}
\hypertarget{hypertarget:Star93}{{Starobinsky}, A.~A. 1993,} \jetpL, 57, 622

\bibitem[{Stevens}, {Scott}, \& {Silk} (1993)]{Stevens93}
\hypertarget{hypertarget:Stevens93}{{Stevens}, D., {Scott}, D., \& {Silk}, J. 1993, \PRL,} 71, 20

\bibitem[{van den Bergh} (2011)]{vdB11Lemaitre}
\hypertarget{hypertarget:vdB11Lemaitre}{{van den Bergh}, S. 2011,} \jrasc, {in press}, \eprint{1106.1195}

\end{thebibliography}
\end{document}